\documentclass[twocolumn,tighten,numberedappendix]{aastex63}
\usepackage{amsmath}

\def\simlt{\lower.5ex\hbox{$\; \buildrel < \over \sim \;$}}
\def\simgt{\lower.5ex\hbox{$\; \buildrel > \over \sim \;$}}

\def\beq{\begin{equation}}
\def\eeq{\end{equation}}
\def\ba{\begin{eqnarray}}
\def\ea{\end{eqnarray}}
\def\bB{{\,\mathbf B}}

\def\Sect{{\rm Section}}

\def\Eq{Equation}
\def\Eqs{Equations}

\def\tauT{\tau_{\rm T}}

\def\sT{\sigma_{\rm T}}

\def\TC{T_{\rm C}}
\def\Teff{T_{\rm eff}}

\def\tC{t_{\rm C}}

\def\vrec{v_{\rm rec}}

\def\Eav{\bar{E}}

\def\gabs{\gamma_{\rm abs}}

\def\E{{\cal E}}

\def\lbar{\lambda\llap {--}}

\def\nph{n_{\rm ph}}

\def\nph{n_{\rm ph}}
\def\tesc{t_{\rm esc}}

\def\l{\ell}
\def\lB{\ell_B}
\def\omRJ{\omega_{\rm RJ}}
\def\epRJ{\epsilon_0}
\def\dns{\dot{n}_{\rm s}}
\def\Teff{T_{\rm eff}}
\def\Theff{\Theta_{\rm eff}}
\def\tesc{t_{\rm esc}}
\def\tdiss{t_{\rm diss}}
\def\ep{\epsilon}

\def\ThC{\Theta_{\rm C}}

\def\Epk{E_{\rm c}}
\def\Qdiss{\dot{U}_{\rm diss}}
\def\QC{\dot{U}_{\rm C}}

\def\s{S}
\def\h{H}
\def\dnph{\dot{n}_{\rm ph}}
\def\npm{n_\pm}
\def\tW{t_{\rm W}}
\def\tdiff{t_{\rm diff}}
\def\Eesc{\overline{E}_{\rm esc}}
\def\epesc{\overline{\ep}_{\rm esc}}

\def\uabs{u_{\rm abs}}
\def\dncs{\dot{n}_{\rm cs}}
\def\fff{q}
\def\dngg{\dot{n}_{\gamma\gamma}}

\newbox\grsign \setbox\grsign=\hbox{$>$} \newdimen\grdimen \grdimen=\ht\grsign
\newbox\simlessbox \newbox\simgreatbox \newbox\simpropbox
\setbox\simgreatbox=\hbox{\raise.5ex\hbox{$>$}\llap
     {\lower.5ex\hbox{$\sim$}}}\ht1=\grdimen\dp1=0pt
\setbox\simlessbox=\hbox{\raise.5ex\hbox{$<$}\llap
     {\lower.5ex\hbox{$\sim$}}}\ht2=\grdimen\dp2=0pt
\setbox\simpropbox=\hbox{\raise.5ex\hbox{$\propto$}\llap
     {\lower.5ex\hbox{$\sim$}}}\ht2=\grdimen\dp2=0pt
\def\simgt{\mathrel{\copy\simgreatbox}}
\def\simlt{\mathrel{\copy\simlessbox}}

\begin{document}

\title{Emission of magnetar bursts and precursors of neutron star mergers} 

\author{Andrei M. Beloborodov}
\affil{Physics Department and Columbia Astrophysics Laboratory, Columbia University, 538  West 120th Street New York, NY 10027}
\affil{Max Planck Institute for Astrophysics, Karl-Schwarzschild-Str. 1, D-85741, Garching, Germany}

\begin{abstract}
Magnetar bursts can be emitted by Alfv\'en waves growing in the outer magnetosphere to nonlinear amplitudes, $\delta B/B\sim 1$, and triggering magnetic reconnection. Similar magnetic flares should occur quasi-periodically in a magnetized neutron star binary nearing merger. In both cases, fast  dissipation in the magnetic flare creates optically thick $e^\pm$ plasma, whose heat capacity is negligible compared with the generated radiation energy. Magnetic dissipation then involves photon viscosity and acts through Compton drag on the plasma bulk motions in the reconnection region. The effective temperature of the resulting Comptonization process is self-regulated to tens of keV. The generated X-ray emission is calculated using time-dependent radiative transfer simulations, which follow the creation of $e^\pm$ pairs and the production, Comptonization, and escape of photons. The simulations show how the dissipation region becomes dressed in an $e^\pm$ coat, and how the escaping spectrum is shaped by radiative transfer through the coat.  The results are compared with observed magnetar bursts, including the recent activity of SGR~1935+2154 accompanied by a fast radio burst. Predictions are made for X-ray precursors of magnetized neutron star mergers.
\end{abstract}



\section{Introduction}

Magnetars are neutron stars with ultrastrong magnetic fields $B=10^{14}$-$10^{16}\,$G (see \cite{Kaspi17} for a review). They display spectacular X-ray activity, including rare giant $\gamma$-ray flares of  luminosities up to $L\sim 10^{47}\,$erg/s and numerous short X-ray bursts with 
$L\sim 10^{39}-10^{42}\,$erg/s. The giant flares are produced in the inner magnetosphere, near the neutron star, and their pulsating tails are emitted by a thermalized fireball confined by the ultrastrong $B\simgt 10^{14}\,$G \citep{Paczynski92,Thompson96}.
The origin of short bursts is not established.

The short bursts have durations of $\sim 0.1\,$s and show rather similar spectra in the broad range of luminosities. They have an exponential cutoff at $\Epk=20-50\,$keV and photon index $\Gamma_{\rm ph}=d\ln N/d\ln E \simgt -1$ at $E<\Epk$ (e.g. \cite{van_der_Horst12,Lin20}). In the absence of a physical emission model, the spectra are usually fitted by phenomenological models --- optically thin thermal bremsstrahlung, double blackbody, or a power-law with an exponential cutoff.

On 2020 April 28, it was discovered that some X-ray bursts of magnetar SGR~1935+2154 are accompanied by emission of a fast radio burst (FRB) \citep{Bochenek20,CHIME20}. In this event, the X-ray burst had a high $\Epk\sim 100\,$keV, and a hard slope $\Gamma_{\rm ph}$, compared with previous bursts from the same source or most bursts observed in other magnetars \citep{Li20,Mereghetti20,Younes20}. The burst had a usual duration $\sim 0.1\,$s and energy output $\E_X\sim 10^{39}\,$erg.

The present paper investigates how the X-ray bursts could be emitted in the outer magnetosphere, where Alfv\'en waves excited by magnetars can grow to nonlinear amplitudes $\delta B/B\simgt 1$ and induce magnetic reconnection events. A magnetohydrodynamic simulation of such events is presented in \cite{Yuan20}. By ``outer'' we mean radii $R\simlt 10^8\,$cm, much larger than the neutron star radius $R_\star=11$-13\,km. In the outer magnetosphere, $B=10^8-10^{11}\,$G is orders of magnitude lower than the surface field $B_\star$. 
We wish to know whether such events can generate 
the X-ray spectrum observed from SGR~1935+2154 activity with FRBs.
We will also investigate how the more typical, softer, magnetar bursts can be produced by magnetic flares in $B=10^8-10^{11}\,$G.

This paper also suggests that the same emission mechanism can continually operate in a neutron star binary nearing merger. 
Interaction of the magnetospheres of two neutron stars 
causes their strong quasi-periodic disturbance $\delta B/B\sim 1$, which triggers magnetic reconnection at $R\sim 10^7\,$cm \citep{Most20}. If the  two stars have surface magnetic fields $B_\star\simgt 10^{12}\,$G, their interaction at $R\sim 10^7\,$cm leads to the same radiative events as in magnetars at $R\simlt 10^8\,$cm. 

The paper is organized as follows. Section~2 describes energy dissipation in the radiative magnetic flares. Then, the emission mechanism is described in Section~3 and simulated numerically in Section~4. The results are discussed in Section~5.


\section{Magnetospheric dissipation}

\subsection{Fast dissipation region}

The typical energy budget of a magnetar burst $\E\simlt 10^{41}\,$erg corresponds to a weak perturbation of the inner magnetosphere $\delta B_\star/B_\star\sim 10^{-3}-10^{-2}$. Such perturbations are expected from starquakes \citep{Blaes89,Thompson96,Bransgrove20}. It is less clear how they generate short X-ray bursts, as this requires quick dissipation of the initially ideal MHD perturbation, on a timescale $\sim 0.1\,$s. 

A possible way for fast dissipation involves launching the perturbation along extended magnetic field lines, into the outer magnetosphere, where the relative wave amplitude $\delta B/B$ grows as $B^{-1/2}\propto R^{3/2}$. Then, the waves can reach $\delta B/B\sim 1$ at $R\sim (30-100)R_\star$ and dissipation can immediately occur through magnetic reconnection, as demonstrated by \cite{Yuan20}. Energy dissipated in such events, $\E_{\rm diss}$, is a fraction of the outer magnetosphere energy, 
\beq
   \E_{\rm mag}(R) \approx 0.1\,\frac{\mu^2}{R^3}= 10^{41}\mu_{33}^2R_8^{-3}\,{\rm erg},
\eeq
where  $\mu=B_\star R_\star^3=10^{33} B_{\star,15} R_{\star,6}^3$ is the magnetic dipole moment of the star. Magnetar bursts have energies consistent with $\E_{\rm diss}\simgt 0.1\E_{\rm mag}$ at $R\sim (0.3-1)\times 10^8\,$cm.

\begin{figure}[t]
 \centering
  \includegraphics[width=0.41\textwidth]{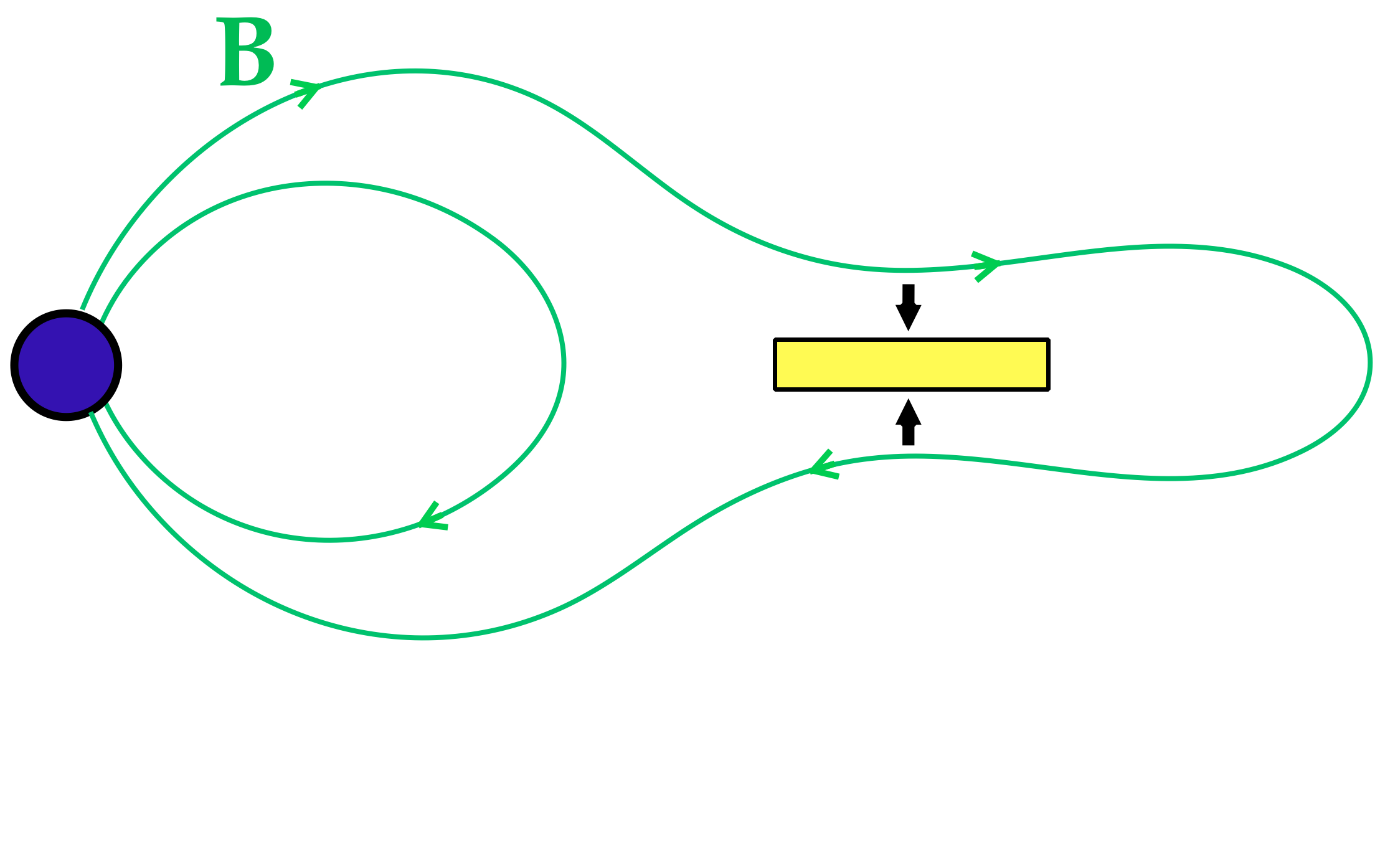}
  \includegraphics[width=0.33\textwidth]{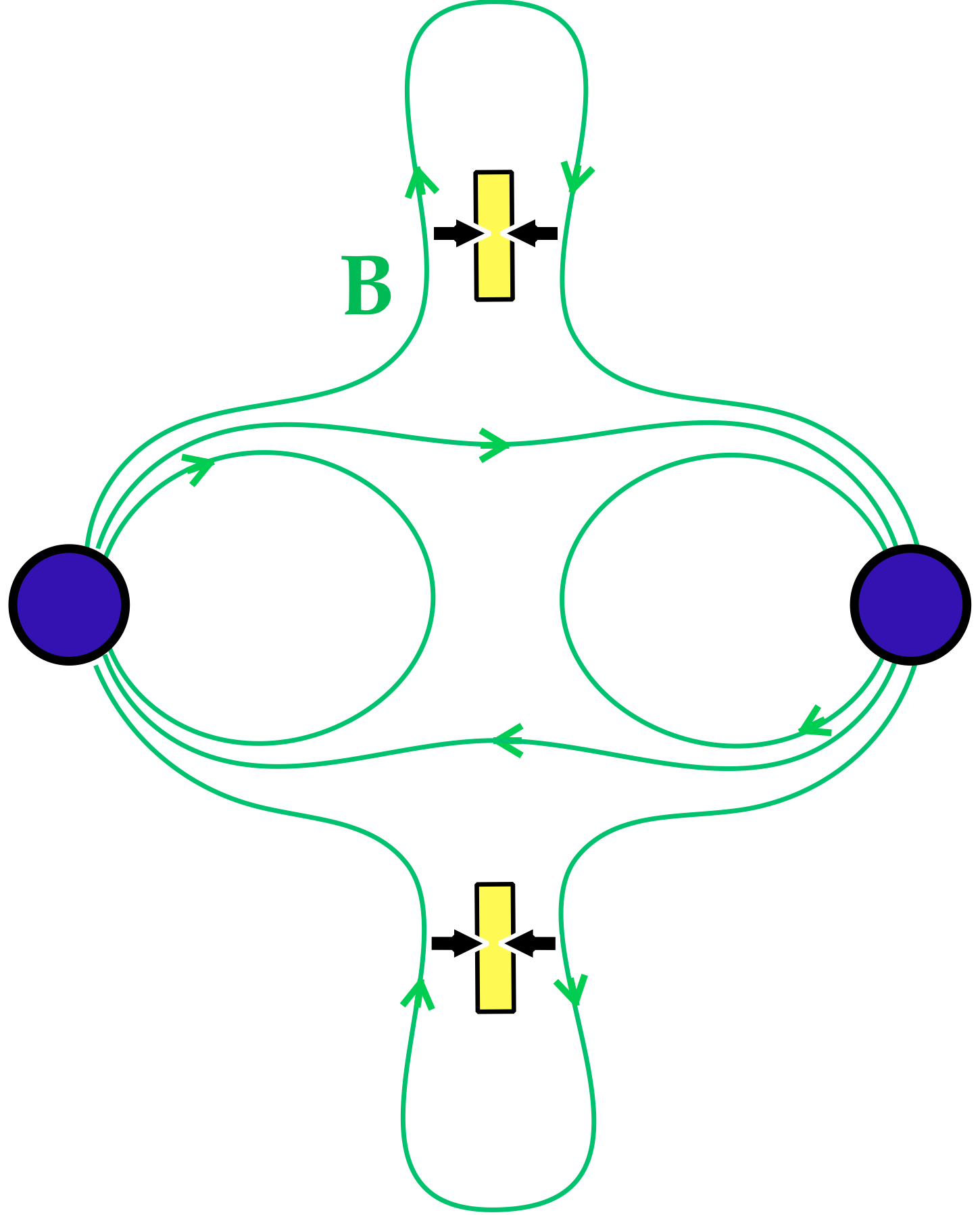}
\caption{
Magnetic flares in active isolated neutron stars (top) and binary neutron stars (bottom).  In both cases, pumping a strong twist into a magnetic loop causes its inflation. Then reconnection is triggered in the yellow regions. Black arrows indicate the inflow of magnetic energy into these regions. Then, the disconnected part of the magnetic loop is ejected, and the remaining part (connected to the star) snaps back to the lower-energy configuration. 
}
\end{figure}

Similar magnetic dissipation events are expected during magnetospheric interaction in a binary neutron star \citep{Most20}. In this case, the field lines connecting the two stars can be continually twisted by their asynchronous rotation, which pumps $\delta B/B\simgt 1$ and drives quasi-periodic magnetic flares.  
Figure~1 shows the flaring magnetosphere of a binary system in the simple case of (anti) parallel magnetic dipole moments of the stars. For comparison,  Figure~1 also illustrates a magnetic flare in an isolated magnetar. 

In both cases, the flare develops through sudden unstable inflation of the magnetosphere outside some radius $R$. It happens where the magnetic twist $\delta B$ exceeds the original field $B$ by a numerical factor $\simgt 1$, which depends on the geometry and the rate of twisting \citep{Parfrey13}.  This inflation creates a configuration with two opposite magnetic fluxes separated by a current sheet. The sheet thickness collapses under the magnetic pressure, and reconnection is triggered by the tearing instability breaking the current sheet into threads.

Relativistic magnetic reconnection occurs with speed $\vrec\approx 0.1c$ \citep{Kagan15}. Therefore, the magnetic flare is expected to occur on the timescale 
\beq
\label{eq:tdiss}
  \tdiss\sim 10\, \frac{R}{c}. 
\eeq
As a result, the inflated magnetic field lines snap back to their original configuration, losing the twist and ejecting plasmoids from the magnetosphere.
The characteristic size of the reconnection region $\s$ (the half-length of the yellow regions in Figure~1) is a fraction of radius $R$. Its half-thickness is $\h\sim (\vrec/c)\s\sim 0.1\s$. 

Part of the magnetic energy flux flowing through the reconnection region $\vrec B^2/8\pi$ gets dissipated. The maximum dissipated fraction of $0.5$ is approached if the guide field $B_{\rm gd}$ (the magnetic field component along the electric current) is small, $B_{\rm gd}\ll \delta B$ (e.g. \cite{Sironi20}). The resulting dissipation rate per unit volume (inside the layer of thickness $2\h=2\s\vrec/c$) may be written as 
\beq
\label{eq:Qdiss}
  \Qdiss=\zeta\,\frac{U_Bc}{\s}, \qquad U_B=\frac{B^2}{8\pi},
\eeq
where $\zeta<0.5$. 

Both analytical theory \citep{Uzdensky10} and first-principle kinetic simulations (e.g. \cite{Sironi14}) show that the current sheet breaks up into a chain of plasmoids with a broad (self-similar) distribution of sizes, from the microscopic Larmor radius to the macroscopic size $\sim \h$. The plasmoids are accelerated by magnetic stresses and ejected along the reconnection layer, allowing reconnection to proceed with speed $\vrec\approx 0.1 c$.

\subsection{Reconnection with photon viscosity}

Reconnection in magnetars and neutron star binaries operates in a highly radiative regime --- essentially all dissipated energy immediately converts to radiation. The heat capacity of the plasma particles is negligible compared with that of photons; particles carry $\sim 10^{-5}$ of the dissipated energy in a typical magnetic flare, as shown below. Under such conditions, reconnection directly heats {\it photons} rather than the plasma. 
Photons receive energy by scattering off the moving plasmoids, and magnetic energy is dissipated by photon viscosity, which damps the plasmoid motions driven by magnetic stresses \citep{Beloborodov17c}. Note also that the fluid bulk motions in radiative reconnection exceed the particle thermal motions measured in the fluid rest frame.

Radiation receives energy as follows:
\\
(1) Photons gain energy in scattering. As long as electron recoil is neglected (Thomson scattering), photons in the reconnection layer continue to drift upward in energy $E$ with a rate proportional to $E$, similar to Fermi acceleration. Their energy gain per Thomson scattering $(\overline{\Delta E}/E)_{\rm T}$ depends on the state of the plasma and defines the effective Comtponization temperature 
\beq
\label{eq:Theff}
    \Theff\equiv \frac{k\Teff}{m_ec^2}
    \equiv \frac{1}{4}\left(\frac{\overline{\Delta E}}{E}\right)_{\rm T}.
\eeq
Compton recoil in scattering is negligible for photons with $E\ll \Teff$, and 
the reconnection layer upscatters such photons just like a static Maxwellian electron gas with temperature $\Teff$ would do. Electron recoil in scattering becomes significant at $E\sim k\Teff$. It steals photon energy $(\overline{\Delta E}/E)_{\rm recoil}=-E/m_ec^2$ and suppresses photon population at $E\gg k\Teff$. The small population of energetic photons with $E\gg k\Teff$ is still crucial, because it controls $e^\pm$ pair creation.
\\
(2) The $e^\pm$ plasma produces new photons with energies $E_0\ll k\Teff$. The rate of photon production $\dnph$ is given in \Sect~3 below. This rate is controlled by the plasma density and its true temperature $T$ measured in the fluid frame, rather than by $\Teff$. 

Photon production gives more photons to share the dissipated power, lowering energy per photon. When emission proceeds in a quasi-steady regime, the average energy of escaping photons $\Eesc$ satisfies the relation
\beq
\label{eq:Eesc}
   \dnph \Eesc\approx \Qdiss,
\eeq
which assumes $\Eesc\gg E_0$. Ratio $\Eesc/k\Teff$ is related to the spectral slope of escaping radiation, which extends from $E_0$ to $k\Teff$.

Creation of $e^\pm$ pairs endows the reconnection layer with a signifcant 
Thomson optical depth, up to $\tauT\sim 10$ in the models calculated below. Then, very large plasmoids can trap and advect radiation while smaller plasmoids move through radiation, experiencing Compton drag. 
This means that photon viscosity operates on scales smaller than the size of the layer. This is similar to photon viscosity in optically thick turbulence cascades \citep{Zrake19}. 

Photon viscosity (Compton drag) implies that current density $J$ in a plasma with density $n_\pm$ is accompanied by energy losses per particle $\dot{E}_e \sim c\sT U(J/en_\pm c)^2$, where $U$ is the radiation energy density. Losses per unit volume, $n_\pm \dot{E}_e$, may be written as $J^2/\tilde{\sigma}$ with effective conductivity $\tilde{\sigma}$. This gives $\tilde{\sigma}\sim c e^2 n_\pm/\sT U$ and the magnetic diffusivity 
\beq
   \eta=\frac{c^2}{4\pi\tilde{\sigma}}\sim c r_e\frac{U}{n_\pm m_ec^2}, \qquad
   r_e\equiv\frac{e^2}{m_ec^2}.
\eeq
A classical Sweet-Parker resistive reconnection layer would have relative thickness $\delta_{\rm SP}/S\sim (\eta/cS)^{1/2}$, and  its tearing instability would create structures as thin as $\delta_\eta\sim 100\eta/c$ \citep{Uzdensky10}. For typical parameters of our burst model $\delta_\eta \ll c/\omega_p$, where $\omega_p=(4\pi e^2 n_\pm/m_e)^{1/2}$ is the plasma frequency. Therefore, the small-scale reconnection does not obey the resistive MHD. Instead, it proceeds in the collisionless regime, with speed $\sim 0.1c$, as confirmed by kinetic simulations with Compton drag \citep{Sironi20}. Note also that the characteristic electron free  path due to photon viscosity may be written as $\lambda\sim S/\l$, where $\l\sim 10^5-10^7$ is the radiation compactness parameter (\Eq~(\ref{eq:l}) below). Our burst models have 
\beq
   S\gg\lambda\gg \frac{c}{\omega_p} \gg \delta_\eta.
\eeq
This is different from reconnection in ultrastrong fields $B>10^{13}\,$G, which could proceed in the resistive MHD regime \citep{Uzdensky11}. 

Despite the strong collisionless dynamics of the reconnection layer on microscopic scales, most of magnetic dissipation occurs on scales $\gg\lambda$, through radiation drag on the moving macroscopic plasmoids. Energy is extracted from the field due to work of the ideal MHD force $\boldsymbol{J}\times\bB/c$ rather than ohmic dissipation of currents.

\subsection{Impulsive  particle acceleration}

The dissipation process described above is similar to pulling bodies (plasmoids) by strings (magnetic field lines) against a viscous background (radiation). Radiative kinetic plasma simulations suggest that this process accounts for $\sim 80$\% of total dissipation \citep{Sironi20}. The remaining fraction goes into nearly impulsive acceleration (``injection'') of nonthermal particles. The injection occurs near X-points in the reconnection layer and operates on the timescale as short as 
\beq
  t_{\rm acc}\sim \frac{B}{4\pi e n_\pm c}=\frac{\omega_B}{\omega_p^2},
\eeq
where $\omega_B=eB/m_ec$ is the gyro-frequency. The maximum accelerating electric field $E\sim (\vrec/c)B\sim 0.1B$ is capable of pushing particles to Lorentz factors $\gamma\sim \sigma_0=B^2/4\pi m_ec^2 n_\pm$ on the timescale $t_{\rm acc}$. However, for the typical parameters of our burst model, radiative losses stop the acceleration at $\gamma_{\rm cr}\sim (S/\l r_e)^{1/4}\sim 10^3<\sigma_0$. Thus, X-point injection is drag-limited, in contrast to previously discussed magnetic flares near accreting black holes, where $\gamma_{\rm cr}>\sigma_0$ \citep{Beloborodov17c}.

The vast majority of injection events give small $\gamma\sim 1$ (see Figure~8 in \cite{Sironi20}); they occur via particle ``pick up'' by outflows from X-points. The simulations with magnetization $\sigma_0=10$ show that a few percent of the dissipated power is spent to inject highly relativistic particles. Note however that magnetic flares around neutron stars have much greater $\sigma_0$, and radiative kinetic simulations with higher magnetizations are needed to clarify the scaling of injection with $\sigma_0$. Below we use a simple parametrization of high-energy particle injection to study its effect on the burst emission.

\subsection{Dissipation of waves with $\delta B/B<1$}

Alfv\'en waves that do not reach $\delta B/B>1$ 
keep bouncing in the closed magnetosphere for some time and eventually dissipate, without triggering large-scale magnetic reconnection. Their dissipation may occur via a turbulence cascade developing due to nonlinear effects \citep{Thompson98,Li19}. Furthermore, dissipation may result from the shearing of Alfv\'en wave packets as they propagate along the curved magnetic field lines, which enhances the electric current density \citep{Bransgrove20,Chen20}. 

The dissipation process is not well understood for the waves with $\delta B<B$. If it does occur quickly, its radiative effect will be similar to that of magnetic reconnection. The radiative calculations below can be applied to any type of fast dissipation in the outer magnetosphere.


\section{Radiative mechanism}
\label{radiation}

\subsection{Parameters of the problem}

\begin{figure}[t]
  \centering
  \includegraphics[width=0.38\textwidth]{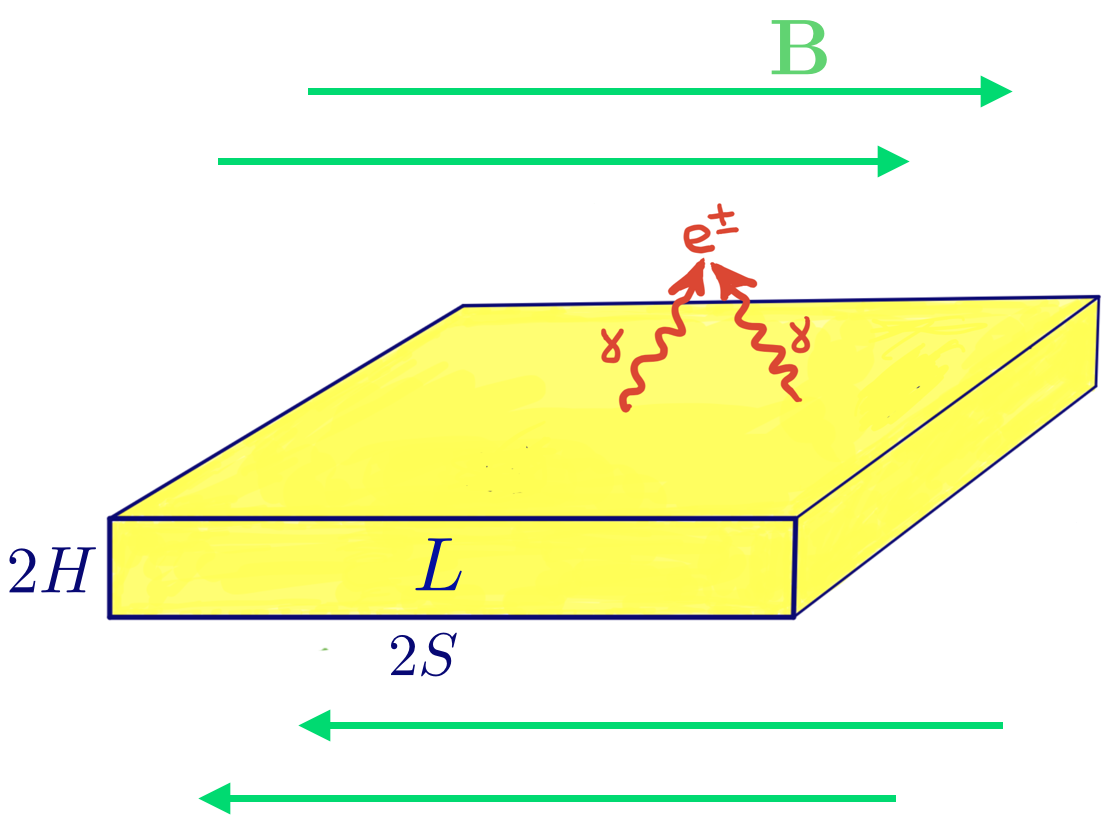}
\caption{Magnetic reconnection releases energy in volume $2\s\times 2\s\times 2\h$ (shaded in yellow). Fraction $\zeta$ of the inflowing field energy dissipates (the remaining fraction flows out through the sides). The released power $L$ converts to radiation, and the region develops an opaque $e^\pm$ coat due to photon-photon collisions. The plasma produces seed (soft) photons, and the escaping spectrum is shaped by their transfer through the $e^\pm$ coat. Photons receive the released energy via Compton scattering.
}
\label{fig:setup}
\end{figure}

The power $L$ dissipated in the magnetic flare immediately converts to radiation, i.e. $L$ is also the radiation production rate. The corresponding dimensionless compactness parameter is defined by
\beq
\label{eq:l}
   \l=\frac{\sT L}{m_ec^3\s}\approx 2.7\times 10^5\, L_{41}\,\s_7^{-1},
\eeq
where $2\s$ is the size of the dissipation region (Figure~\ref{fig:setup}).
The flare is fed by magnetic energy in the dissipation region of volume $V\sim  2\s\times 2\s \times 2\h$.\footnote{The dissipation region may be composed of $N$ regions of volume $2\s\times 2\s\times 2\h$, giving the total observed luminosity $L_{\rm tot}=NL$. For example, reconnection in an axisymmetric magnetosphere forms a  dissipation region $2\pi R\times 2\s\times 2\h$; then $N\sim 10$.} The volume-average energy release rate is
\beq
\label{eq:Qdiss1}
  \Qdiss\approx\frac{L}{V}\approx \frac{m_ec^3\,\l}{8\s\h\sT}. 
\eeq
It may also be expressed as in \Eq~(\ref{eq:Qdiss}), which gives the relation
\beq
\label{eq:l_lB}
  \frac{\l}{\lB}\approx \left(\frac{\h}{0.1\s}\right)\,\zeta,
\eeq
where
\beq
\label{eq:lB}
   \lB\equiv \frac{\sT U_B\s}{m_ec^2} \approx 3.2\times 10^5\,\s_7 B_9^2.
\eeq
The only dimensional parameter of the problem is $\s\sim 10^7\,$cm; it determines $B$ for given $\l$ and $\zeta$.

The formulation of the burst problem is simple:\footnote{This formulation of the burst problem is applicable to both magnetic reconnection and dissipation of Alfv\'en wave turbulence.}
there is an initially empty region of volume $V$ where magnetic energy is released with a given power $L$. The dissipation event lasts for $\tdiss\gg \s/c$ (\Eq~\ref{eq:tdiss}). The magnetic field $B$ in the region  and the released power $L$ are related by \Eq~(\ref{eq:l_lB}). The question is what radiation is emitted by this dissipation event.

In a real magnetosphere, the ``empty'' region contains some seed plasma and radiation. Magnetic dissipation starts with accelerating seed particles, producing inverse Compton emission and igniting pair creation.
Regardless of the initial state, the high compactness parameter $\l$ guarantees that the region becomes filled with dense radiation and $e^\pm$ plasma 
\citep{Cavallo78}. 

The $e^\pm$ density $\npm$ in the dissipation region is not known in advance and needs to be calculated self-consistently. It will determine the characteristic optical depth 
\beq
   \tauT\sim \sT \h \npm.
\eeq
A simple estimate shows that the self-consistent $n_\pm$ is much lower than the photon number density $\nph=U/\Eav$, where $\Eav$ is the average photon energy inside the source and $U$ is the radiation density. When the burst is emitted via quasi-steady diffusion of radiation, one can estimate $L\sim VU/\tdiff$, where $\tdiff\sim \tauT\h/c$. This gives the relation 
$L\sim 8\s^2\Eav \nph c/\sT\h n_\pm$, and so,
\beq
\label{eq:npm}
  \frac{n_\pm}{\nph}\sim  8\,\frac{\s}{\h}\,\frac{\Eav}{m_ec^2}\;\l^{-1}\ll 1.
\eeq
This also gives an estimate for the magnetization parameter of the dissipation region,
\beq
\label{eq:sigma0}
   \sigma_0=\frac{B^2}{4\pi n_\pm m_ec^2}\sim2\,\frac{\h}{\s}\,\frac{\lB}{\tauT}.
\eeq

\subsection{Comptonization}

We will use a uniform statistical description of the plasma motions in the dissipation region. Photons have macroscopic free paths $\sim\h/\tauT$ and randomly sample the four-velocity $u=\gamma\beta$ of the scattering particles. In our simulations, $u$ in each scattering event is drawn from a single global particle distribution $f_e(u)$. This approximation is much simpler than a full kinetic plasma simulation of magnetic reconnection coupled to radiative transfer, a challenge for future work. 

The main parameter of $f_e(u)$ is the integral
\beq
\label{eq:Theff1}
   \Theff=\frac{k\Teff}{m_ec^2}=\frac{1}{3}\int u^2 f_e(u) du.
\eeq 
It determines the rate of energy gain of radiation from scattering $\QC$ (see the definition of $\Teff$ in \Eq~\ref{eq:Theff}). Note that $\Theff$ is not known in advance and can evolve in the self-consistent radiative model of the burst. For any given instantaneous state of radiation, $\Teff(t,x,y,z)$ must satisfy the local energy balance,
\beq
\label{eq:Cbalance}
   \QC(\Teff)= \Qdiss.
\eeq
It states that the plasma immediately passes the dissipated energy to radiation, as the plasma heat capacity is negligible. The energy exchange rate $\QC(\Teff)$ depends on both the spectrum of radiation and the shape of $e^\pm$ distribution, $f_e(u)$. In our Monte-Carlo simulations below, $\QC(\Teff)$ is calculated numerically using the exact Klein-Nishina cross section for Compton scattering. 

A simple analytical estimate for $\QC(\Teff)$ is obtained using Thomson approximation and assuming isotropic $e^\pm$ motions with $\Theff\ll 1$:
\beq
\label{eq:QC_Th}
    \QC\approx 4c\sT \npm U \left(\Theff-\ThC\right), \quad
     \ThC\approx \frac{\int \ep U_\ep d\ep}
    {4U}.
\eeq
Here $U_\ep$ is the spectral energy density of radiation, $U=\int U_\ep\,d\ep$ is the total energy density, and $\ep= E/m_ec^2$.
In a quasi-steady state, $U\sim \Qdiss\tdiff$ where $\tdiff\approx \tauT \h/c$ is the diffusion time of radiation out of the heating region. Then the energy balance condition~(\ref{eq:Cbalance}) simplifies to 
\beq
\label{eq:CbalanceT}
    4\tauT^2(\Theff-\ThC)\sim 1.
\eeq

The parameter $\Theff=\langle u^2/3\rangle$ can be determined for any given distribution function $f_e(u)$, which may have different shapes. In a standard model of thermal Comptonization, $f_e(u)$ would be Maxwellian,
\beq
\label{eq:M}
   f_e(u) \propto  u^2 e^{-\gamma/\Theff},
   \quad 
\eeq
where $\Theff$ is the only parameter. A realistic reconnection region is not quite Maxwellian:  $f_e(u)$ has a sub-relativistic peak formed by plasmoid motions and a high-energy tail formed by impulsive particle acceleration near X-points. The tail steeply declines at $u\gg 1$. A simple example of a non-Maxwellian distribution is  
\beq
\label{eq:fe}
   f_e(u)\propto \frac{u^2}{1+(u/u_0)^{2+\alpha}}.
\eeq
It replaces the Maxwellian exponential cutoff with a power-law decline at  $u\gg u_0$. The distribution function in \Eq~(\ref{eq:fe}) has two parameters $u_0$ and $\alpha$. $\Theff(u_0,\alpha)$ can be calculated from \Eq~(\ref{eq:Theff1}), and a good approximation is given by 
\beq
  \Theff\approx \frac{\alpha-1}{5(\alpha-3)}\,u_0^2.
\eeq
The power received and radiated by the high-energy tail at $u\gg u_0$ is proportional to $\int_u^\infty  {u^\prime}^2 f_e(u^\prime)\,du^\prime \propto u^{-\alpha+3}$. In particular, the fraction of the dissipated power received by particles with $u>1$ is approximately given by 
\beq
\label{eq:U1}
     \frac{\dot{U}_1}{\Qdiss}\approx \frac{5}{\alpha+2}\, u_0^{\alpha-3}.
\eeq
In magnetic flares, the power deposited into relativistic particles $u\gg 1$ is likely a small fraction of the total released power $\Qdiss$. This corresponds to $f_e(u)$ with a steep high-energy tail, $\alpha>5$. A strong nonthermal tail, $\alpha=5$, gives $\dot{U}_1/\Qdiss\approx 2\Theff$, which varies around 5\% in the sample models calculated below.

The effect of the nonthermal high-energy tail in $f_e(u)$ will be illustrated below by calculating the burst spectra for the two choices of  $f_e(u)$, \Eqs~(\ref{eq:M}) and (\ref{eq:fe}).
In both Maxwellian and nonthermal plasma models, $f_e(u)$ has a single-peak shape: it grows as $f_e(u)\propto u^2$ at small $u^2<\Theff$ and falls off at high $u$. Therefore, the production of Comptonized radiation declines at $E\gg k\Teff$. The weak high-energy part of the photon spectrum is important, because it controls the rate of $e^\pm$ creation in photon-photon collisions $\dngg$ (see \Sect~\ref{pairs} below). However, the feedback of $\dngg$ on $\Teff$ turns out modest, leading to $k\Teff=10-40\,$keV for the entire range of relevant parameters (\Sect~\ref{Teff}).

\subsection{Photon production processes}
\label{photons}

The single global distribution $f_e(u)$ is convenient in the calculations of Comptonization. However, for the analysis of photon production, it is more convenient to view the plasma as the sum of two parts: thermal particles (which have a Maxwellian distribution in the plasma rest frame) and nonthermal particles injected near X-points.

The true plasma temperature $T$ (which enters the photon production rates given below) differs from $\Teff$.  It is defined in the plasma rest frame and locked to the Compton temperature,
\beq
    T=\TC.
\eeq
This equality is enforced on the Compton timescale $\tC$, which is much shorter than the light crossing time $\s/c$,
\beq
   \frac{c\tC}{\s}\approx \frac{3m_ec^2}{8 \sT\s \,U}=\frac{3}{8\lB} \frac{U_B}{U}\ll 1.
\eeq
The value of Compton temperature is found from 
\beq
\label{eq:QC0}
   \QC(\TC)=0,
\eeq
where $\QC(T)$ is calculated for a Maxwellian plasma (in its rest frame).
A simple estimate for $\ThC\equiv k\TC/m_ec^2$ is obtained in the Thomson approximation and given in \Eq~(\ref{eq:QC_Th}). An accurate calculation of $\TC$ would use the radiation spectrum $U_\ep$ Doppler boosted to the plasma rest frame. However, only a small fraction of the reconnection region (small plasmoids) has highly relativistic bulk speeds, and the thermal photon production is dominated by large plasmoids with mildly relativistic motions. Therefore, we will neglect the Doppler boost and find $\TC$ from condition~(\ref{eq:QC0}) in the fixed lab frame; this gives $\sim 10\%$ accuracy of $T$, leading to a similar accuracy of $\dnph$. 

In general, thermal plasma in compact X-ray sources produces photons through three main processes: bremsstrahlung, double Compton scattering, and cyclo-synchrotron emission. These processes have been well studied, in particular in the context of accreting black holes and GRB explosions  \citep{Svensson84,Wardzinski00,Beloborodov13,Vurm13,Thompson14}. 
The cyclo-synchrotron mechanism dominates in the magnetic flares of interest here. In addition, some photons may be produced by synchrotron emission from high-energy particles. The photon production mechanisms are summarized below.

\subsubsection{Cyclo-synchrotron emission}

 The thermal cyclo-synchrotron emission peaks at high harmonics $\fff\,\omega_B$ of the electron gyro-frequency $\omega_B=eB/m_ec$, with $\fff\approx 15-20$. 
The emission at frequencies $\omega<\fff\,\omega_B\equiv \omRJ$ is self-absorbed, forming a Rayleigh-Jeans spectrum, and at $\omega>\omRJ$ Compton upscattering  wins over absorption. At $\omega=\omRJ$ the two processes occur with equal rates,
\beq
\label{eq:abs}
   \mu_{\rm abs} (\omRJ)=  4\Theff \sT \npm,
\eeq
where $ \mu_{\rm abs}(\omega)$ is the absorption coefficient of the thermal plasma.

The self-absorbed Rayleigh-Jeans spectrum peaks at  $\omega=\omRJ$ and
has the photon density 
\beq
\label{eq:RJ}
   n_{\rm RJ}\approx \frac{\Theta\,\epRJ^2}{2\pi^2\lbar^3},
   \qquad \epRJ\equiv\frac{\hbar\omRJ}{m_ec^2}=\fff\,\frac{B}{B_Q},
\eeq
where $\lbar=\hbar/m_ec$ is Compton wavelength, and $B_Q= m_e^2c^3/\hbar e\approx4.44\times 10^{13}\,$G.
Comptonization moves photons from $\omRJ$ to higher frequencies on the timescale $t_{\rm IC}=(4\Theff c \sT \npm)^{-1}$, and the photons with $\omega\sim\omRJ$ are resupplied by the cyclo-synchrotron emission with the rate 
\beq
  \dncs \sim \frac{n_{\rm RJ}}{t_{\rm IC}}. 
 \eeq
 The numerical coefficient in this relation is close to unity and may be refined
as $3/4$ using Kompaneets equation \citep{Vurm13}. This gives
\beq
\label{eq:dncs}
   \dncs 
   \approx \frac{3 \alpha_f \sT}{2\pi^2 m_ec}\,\npm\,\Theta\Theff B^2  \fff^2,
\eeq
where $\alpha_f=e^2/\hbar c\approx 1/137$ is the fine structure constant.

The exact $\fff=\omRJ/\omega_B$ depends on $B$ and $\Theta$, and involves cumbersome calculations \citep{Wardzinski00}. However, its value falls in a narrow range, because $\omRJ$ is far in the exponentially steep tail of the thermal cyclo-synchrotron emissivity, $\fff\gg 1$. A simple approximate fit suggested by \cite{Vurm13} for a range of $B$ and $\Theta$ similar to what we need below is 
\beq
\label{eq:fff}
    \fff\equiv \frac{\omRJ}{\omega_B}
    \approx 17 \,B_9^{-0.1}\left(\frac{\Theta}{0.05}\right)^{0.3}.
\eeq
The weaker dependence of $q$ on $\Theff$ is omitted here (and the simulations below show that $\Theff$ is not far above $\Theta$). \Eq~(\ref{eq:fff}) gives a reasonable approximation, with accuracy of a few tens of percent.

\subsubsection{Double Compton scattering and bremsstrahlung}

The thermal $e^\pm$ plasma also produces photons through bremsstrahlung (B) and double Compton scattering (DC), with the following rates (e.g. Svensson 1984),
\beq
  \dot{n}_{\rm B}\sim 0.1 c \sT n^2 \Theta^{-1/2}, \quad
 \dot{n}_{\rm DC}\sim 0.1 c\sT n n_{\rm ph} \Theta^2.
\eeq
The ratio of the cyclo-synchrotron photon production rate to the DC scattering rate is 
\beq
\nonumber
    \frac{\dncs}{\dot{n}_{\rm DC}}
    \sim \frac{3\, \epRJ^2\Theff}{2\,\nph \lbar^3\Theta} \\
    \approx 10^2 \left(\frac{\fff}{20}\right)^2 \frac{\bar{E}}{m_ec^2}\,\frac{U_B}{U}\,\frac{\Theff}{\Theta}, 
\eeq
where $\bar{E}$ is the average photon energy in the radiation spectrum.
Our calculations below give $\bar{E}\simgt 0.03 m_ec^2$, and we find $\dncs>\dot{n}_{\rm DC}>\dot{n}_{\rm B}$.

\subsubsection{Synchrotron photons from  high-energy particles}

An additional source of photons is the synchrotron emission from nonthermal particles. Particles with a Lorentz factor $\gamma$ emit synchrotron photons with the characteristic energy 
\beq
   \ep_s=\frac{E_s}{m_ec^2}\sim 0.3\,\gamma^2\frac{B}{B_Q}.
\eeq
Here, the choice of the numerical coefficient $\sim 0.3$ corresponds to isotropic particles, which may be a rough approximation, however it will be sufficient for the estimates below. Absorption by the thermal plasma imposes a lower limit $\ep_s>\epRJ$ (\Eq~\ref{eq:RJ}) for photons that survive and engage in the Comptonization process. This lower limit corresponds to 
\beq
    \gamma>\gabs\sim (3\fff)^{1/2}\sim 7.
\eeq
The particles lose energy to synchrotron emission with rate $\dot{\gamma}m_ec^2=-(4/3)c\sT U_B\gamma^2\beta^2$, producing photons with rate $\dot{\gamma}/\ep_s$. This gives the photon production rate per unit volume 
\beq
\label{eq:dns}
    \dns\sim \frac{4\sT U_B}{m_ec}\,\frac{B_Q}{B}\,\npm\int_{\uabs}^\infty f_e(u)\,du,
\eeq
where $\uabs=(\gabs^2-1)^{1/2}\approx\gabs$ is the four-velocity that corresponds to $\gabs$. 

It is useful to relate $\dns$ to the rate of energy deposition into particles with $\gamma>\gabs$, which we denote $\dot{U}_{\gamma>\gabs}$. Self-absorption by nonthermal particles of interest is negligible, and a large fraction of $\dot{U}_{\gamma>\gabs}$ is radiated via synchrotron emission rather than Compton scattering. Therefore, 
\beq
   \dot{U}_{\gamma>\gabs}\sim \frac{4}{3}\,c\sT U_B \npm \int_{\uabs}^\infty u^2 f_e(u) du.
\eeq
Estimating $\int_{\uabs}^\infty u^2 f_e(u) du\sim 2\uabs^2 \int_{\uabs}^\infty f_e(u) du$, we find
\beq
\label{eq:dns1}
    \dns\sim \frac{1}{2q}\,\frac{B_Q}{B}\,\frac{\dot{U}_{\gamma>\gabs}}{m_ec^2}.
\eeq
It should be compared with the required photon production in the self-regulated emission $\dnph\sim \Qdiss/\Eesc$ (\Eq~\ref{eq:Eesc}). One can see that $\dns\ll\dnph$ as long as 
\beq
  \frac{\dot{U}_{\gamma>\gabs}}{\Qdiss}\ll \frac{2q}{\epesc}\,\frac{B}{B_Q}\sim 10^{-2}\,B_9\,\left(\frac{\epesc}{0.1}\right)^{-1},
\eeq
where $\epesc=\Eesc/m_ec^2$. This condition can be easily satisfied in strong $B\simgt 10^9\,$G.

\subsection{Photon balance}
\label{Teff}

In general, the boundary $\epRJ$ between the Rayleigh-Jeans and the Comptonization part of the photon spectrum determines $n_{\rm RJ}=\epRJ^2 \Theta / 2\pi^2\lbar^3$ and implies photon supply with rate $\dnph\approx 3n_{\rm RJ}\Theff c\sT\npm$. This gives the relation
\beq
\label{eq:photons1}
    c\sT\npm \ep_0^2 \Theta\Theff\approx 2\pi \lbar^3\dnph.
\eeq
It is valid regardless of the photon emission mechanism. In a quasi-steady state, the condition $\dnph\approx\Qdiss/\Eesc$ gives
\beq
\label{eq:photons2}
    \sT\npm\approx \frac{2\pi \lbar^3\Qdiss}{c\Theta\Theff\Eesc\ep_0^2}.
\eeq
This ``photon balance'' condition imposes a relation between $n_\pm$ and $\Theff$.

Photon production in the neutron star bursts is dominated by cyclo-synchrotron emission, which has a special feature: $\ep_0^2$  is proportional to the magnetic energy density $U_B$. Since $\Qdiss=\zeta U_Bc/\s$ also scales with $U_B$, the parameter $\lB$ practically drops out from \Eq~(\ref{eq:photons2}), and one finds 
\beq
\label{eq:npm1}
   \sT\npm\s \approx \frac{\zeta}{4\alpha_f \fff^2 \Theta\Theff \epesc}.
\eeq
Note that $\Theta$, $\Theff$, and $\epesc$ are all comparable. The detailed simulations presented below show that $\Theta$ is below $\Theff$, and $\epesc$ is above, so that $\Theta\Theff\epesc\sim \Theff^3$. Then, \Eq~(\ref{eq:npm1}) yields
\begin{eqnarray}
\nonumber
   \Theff & \approx & \left(\frac{\zeta\,\h/\s}{4\alpha_f \fff^2 \tauT}\right)^{1/3} \\
   & \approx & 0.06 \left(\frac{\tauT}{5}\right)^{-1/3} 
                      \left(\frac{\fff}{17}\right)^{-2/3}  \left(\frac{\l}{0.1\lB}\right)^{1/3},
 \label{eq:T-tau}
\end{eqnarray}
where we substituted $\zeta\h/\s\approx 0.1\l/\lB$ (\Eq~\ref{eq:l_lB}).

\subsection{Pair creation and annihilation}
\label{pairs}

The evolution of $e^\pm$ density $n_\pm$ obeys the equation
\beq
  \frac{dn_\pm}{dt}=\dngg-\dot{n}_{\rm ann}, \qquad
    \dot{n}_{\rm ann}\approx\frac{3}{16}\,\sT c \npm^2.
\eeq
Here $\dot{n}_{\rm ann}$ is the annihilation rate, written in the limit of $\Theta\ll 1$ (relativistic corrections to $\dot{n}_{\rm ann}$ are negligible at temperatures of interest $\Theta\simlt 0.1$, see \Eq~(68) in \cite{Svensson82}).
Pairs are created with rate $\dngg$ in collisions of photons with energies above $m_ec^2$, which receive a small fraction of the dissipated power.

The rate $\dngg$ depends on the radiation spectrum formed by Comptonization. Radiation Comptonized by a Maxwellian distribution with temperature $\Teff$ at high energies approaches a Wien spectrum,
\beq 
    U_\epsilon \approx \frac{U_{\rm W}\, \epsilon^3}{6\Theff^4} 
    \exp\left(-\frac{\epsilon}{\Theff}\right)
  \qquad   (\epsilon\gg \Theff),
\eeq 
with the effective $U_{\rm W}\sim (0.1-0.3) U$ depending
on the overall  shape of the Comptonized spectrum with the total density $U$.  
Pair creation in the Wien tail is given by \citep{Svensson84}
\beq
   \dngg\approx 0.02\,\frac{\sT U_{\rm W}^2}{m_e^2c^3 \Theff^5} \exp\left(-\frac{2}{\Theff}\right).
\eeq
When annihilation balance is established, $\dngg\approx\dot{n}_{\rm ann}$, the $e^\pm$ density becomes
\beq
\label{eq:npmW}
  n_\pm \sim \frac{0.1 U}{m_ec^2}\, \Theff^{-5/2} \exp\left(-\frac{1}{\Theff}\right).
\eeq
Then the relation $n_\pm m_ec^2/U\sim 8 \s/\h\l$ (\Eq~\ref{eq:npm}) gives
\beq
\label{eq:Teff_M}
   \Theff^{5/2} \exp\left(\frac{1}{\Theff}\right)\sim 10^{-2}\frac{\h}{\s}\,\l \gg 1.
\eeq
This requires $\Theff\approx 0.05-0.1$ in the entire relevant range of compactness parameter $\l$, and implies that $\Theff$ very slowly decreases with $\l$.
The rough estimate of the numerical coefficient in \Eq~(\ref{eq:Teff_M}) is sufficient, because its variation weakly affects the solution for $\Theff$.

Comptonization by a nonthermal $e^\pm$ distribution $f_e(u)$ (\Eq~\ref{eq:fe}) produces a radiation spectrum with a power-law (rather than exponential) tail at $E\gg k\Teff$. 
This leads to more efficient pair creation, so that the same $\tauT$ can be achieved at a lower $\Teff$.

\subsection{Comptonization timescale}

One can now check how quickly the injected photons with $E_0\ll k\Teff$ get upscattered to energy $E\sim 3k\Teff$ (where losses to Compton recoil become dominant).  Photons gain energy $\Delta E/E=4\Theff$ per scattering and reach the peak after $N_{\rm W}\approx (4\Theff)^{-1}\ln (3k\Teff/E_0)$ scatterings. This takes time $\tW =N_{\rm W}/c\sT\npm$, and so
\beq
   \tW\approx \frac{\ln(3\Theff/\ep_0)}{4 c\sT n_\pm \Theff}.
\eeq
The regime of saturated Comptonization occurs if $\tW$ is much shorter than the timescale for photon escape from the dissipation region, $\tesc$. The photons can escape by diffusing across the reconnection layer on the timescale $t_{\rm diff}\approx \tauT \h/c$, and
\beq
 \frac{t_{\rm diff}}{\tW} \approx \frac{4\Theff \tauT^2}{\ln(3\Theff/\ep_0)}.
\eeq 
As long as $\tdiff$ represents the characteristic residence time of photons in the dissipation region, the ratio $\tdiff/\tW$ controls the overall shape of the escaping radiation spectrum \citep{Illarionov72}. The limit of $\tdiff/\tW\rightarrow \infty$ gives the Wien spectrum $d\nph/d\ln E=0.5\, (E/kT)^3\exp(-E/kT)$, which peaks at $E=3kT$. A small $\tdiff/\tW\simlt 1$ gives a soft spectrum with an exponential cutoff at $E\sim k\Teff$. 

Using the photon balance condition (\Eq~\ref{eq:T-tau}), we find
\beq
\label{eq:tW_tdiff}
     \frac{t_{\rm diff}}{\tW} \sim 1\,\left(\frac{\Theff}{0.06}\right)^{-5}
     \left(\frac{\l}{0.1\lB}\right)^{2}\left(\frac{q}{17}\right)^{-4}.
\eeq
Since $\Theff$ is self-regulated, so that it varies in a narrow range, \Eq~(\ref{eq:tW_tdiff}) gives $0.3\simlt \tdiff/\tW \simlt 3$ in a broad region of the parameter space. The variation of $\tW/\tdiff$ between different models presented below causes (moderate) variations in the spectral slopes. In all cases, Comptonization occurs in the unsaturated regime.

\subsection{Magnetic flares with $\tauT>c/\vrec$}
\label{advection}

In a reconnection layer, the moving magnetic field lines advect the $e^\pm$-photon fluid (coupled by scattering). If $\tauT$ exceeds $c/v_{\rm rec}$, radiation diffusion becomes slower than advection. Then a large part of the generated radiation can be advected sideways along the reconnection layer and ejected together with the magnetic plasmoids on the timescale $\sim \s/c$. This timescale is shorter than $\tdiff$ when $\tauT>c/\vrec$.

In general, the photon residence time in the quasi-steady dissipation region is $\tesc\sim\tdiff$ in the diffusion-dominated regime, and $\tesc\sim\s/c$ in the advection-dominated regime:
\begin{eqnarray}
   t_{\rm esc}\approx \left\{\begin{array}{lr}
   \tauT \h/c & \quad \tauT\simlt  c/\vrec \\
   \s/c   & \quad \tauT\simgt c/\vrec
                                         \end{array} \right.
\end{eqnarray}
The high $\tauT$ in the advection-dominated regime must be accompanied by a reduced $\Theff$, according to the photon balance condition (\Eq~\ref{eq:T-tau}). This situation occurs if the dissipation generates a nonthermal $e^\pm$ distribution with a significant high-energy tail. Compared with a Maxwellian plasma, the nonthermal $e^\pm$ are more efficient in Comptonizing photons to $E>m_ec^2$, enhancing $e^\pm$ creation, so that a high $\tauT>10$ can be achieved even when $\Theff$ is below 0.05 (a numerical example will be calculated in \Sect~4).

In the advection dominated regime, the ratio $\tesc/\tW$, which controls the slope of the Comptonized spectrum, becomes
\beq
     \frac{t_{\rm esc}} {\tW}\sim 1 \,\left(\frac{\Theff}{0.04}\right)^{-2}
     \left(\frac{\l}{0.1\lB}\right)^{1}\left(\frac{q}{17}\right)^{-2}.
\eeq


\section{Radiative transfer simulations}

We have developed a new radiative transfer code \texttt{CompPair}  to calculate the production of Comptonized radiation by $e^\pm$ plasma that is  self-consistently created in photon-photon collisions. The code follows the evolution of radiation and plasma in time. It employs a Monte-Carlo technique to track the emission, scattering, absorption, and escape of a large number of photons. The Monte-Carlo method is combined with a grid-based description of radiation in  phase space, which is used  in the calculation of photon-photon absorption opacity. Monte-Carlo photons in the simulation carry dynamic weights, changed by photon splitting and merging, as needed for sufficient sampling across the radiation spectrum. This allows the code to resolve the weak spectral tail $E\simgt m_ec^2$ that is responsible for $e^\pm$ production. The simulation also follows photons emitted by $e^\pm$ annihilation, which contribute to radiation at $E\simgt m_ec^2$. The code has been well tested.\footnote{One test has a particularly useful setup: energy $\E$ is injected in a closed box (with reflecting boundaries), with no subsequent emission of soft photons, $\dnph=0$. This closed system conserves both energy and the total number of photons and $e^\pm$, $N=N_{\rm ph}+N_\pm$. It relaxes to a steady state with a Wien radiation spectrum and an equilibrium $N_\pm$, reproducing the known analytical result \citep{Svensson84}. This test is not passed unless the scattering, photon-photon collisions, and annihilation are all calculated exactly, with relativistic corrections.}
Further details of the code will be described elsewhere. 

The simulations presented in this paper have $\sim 10^7$ photons in the computational box at any given time. The photons are produced with the rate $\dnph(n_\pm,T)$ given in \Sect~3, and escape when they reach the box boundaries. The box is Cartesian, $2\s\times 2\s \times 2H_{\rm box}$. It has the horizontal half-width $\s=10^7\,$cm and the half-height $H_{\rm box}=0.4\s$. This height is sufficient, as we find that the scattering photosphere is located at smaller altitudes. The evolution of $e^\pm$ plasma parameters $n_\pm$, $T$, and $\Teff$ is calculated on a spatial grid. The problem is symmetric about the midplane $z=0$, and the grid has $N_z=20$ points in the vertical direction for $0<z<H_{\rm box}$. We have checked that this grid gives a reasonably good resolution, as the grid cells have optical depths well below unity.

The energy release rate $\Qdiss$ is prescribed with a Gaussian profile, which peaks at $z=0$,
\beq
\label{eq:heating}
   \Qdiss(t,z)=\dot{U}_0(t)\exp\left(-\frac{z^2}{2\h^2}\right). 
\eeq
During an initial short time interval ($0<t<0.2\s/c$), $\dot{U}_0$ is linearly increased from zero  to a maximum $\dot{U}_0^{\max}$. Then, $\dot{U}_0=\dot{U}_0^{\max}$ is kept constant for a much longer time $t\sim 30\s/c$. 
The released power  in the box is given by
\beq
  L=(2\s)^2\int_{-H_{\rm box}}^{H_{\rm box}} \Qdiss\,dz
  \approx 10\,\s^2\h\dot{U}_0.
\eeq
We parameterize the heating rate using the compactness parameter $\l$,
\beq
\dot{U}_0^{\max}=\frac{m_ec^3\,\l}{10\h\s\sT}=\zeta\,\frac{U_Bc}{\s}.
\eeq
All the runs presented below have $\h=0.1\s$. In this case, $\l=\zeta\lB$.

We have performed two different sets of  simulations, for the two versions of the particle distribution ansatz $f_e(u)$: a Maxwellian (\Eq~\ref{eq:M}) and a smoothly broken power law (\Eq~\ref{eq:fe}). 

The choice of an initial state of radiation and $e^\pm$ plasma is not important, because the system forgets it at times $t\gg\s/c$ and approaches a quasi-steady state. This later phase dominates the observed emission, as the burst duration is $\gg\s/c$ (\Sect~2.1). We have run the numerical models for long times $t\sim 30\s/c$ and here show radiation and $e^\pm$ plasma created during the quasi-steady state, well after the initial relaxation phase. At these times, the balance between pair creation and annihilation is approached even at high altitudes $z$, where the $e^\pm$ density is low.
The presented simulations use the approximate photon production rate $\dnph\approx\dncs$ given by \Eqs~(\ref{eq:dncs}) and (\ref{eq:fff}); the  photons are injected with a Planck distribution with the average energy $E_0=\hbar\omRJ$. Our test runs have shown that using more accurate modules for soft photon injection weakly affects the results; they will be needed in future simulations where an accurate particle distribution function is provided by a kinetic plasma code coupled to radiative transfer.

\begin{figure}[t]
    \centering
  \includegraphics[width=0.64\textwidth]{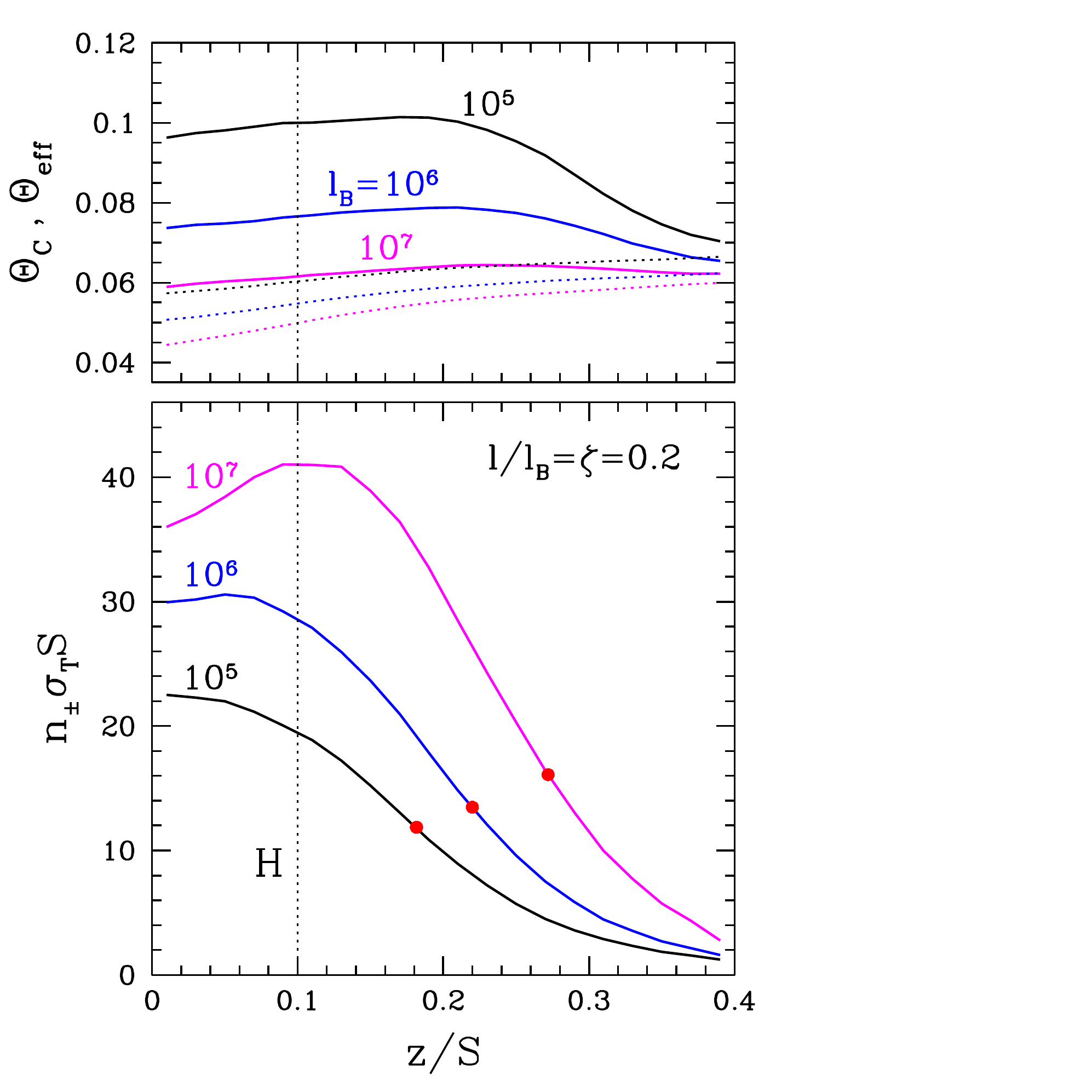}
\caption{Structure of the $e^\pm$ coat formed in three simulations ($\lB=10^5$, $10^6$, and $10^7$) with a Maxwellian plasma. All three simulations had the same $\zeta=0.2$, and so their compactness parameters are $\l=0.2\lB$. Bottom: the $e^\pm$ density profile $n_\pm(z)$. The red circle on each curve marks the position of the photosphere --- the location where the optical depth remaining to escape along the $z$ axis is unity. Top: the plasma temperature $\Theff(z)$ (solid) and the radiation Compton temperature $\ThC(z)$ (dotted). The vertical dotted line indicates the characteristic altitude $z=H$ above which the heating $\Qdiss$ is exponentially suppressed (\Eq~\ref{eq:heating}).
}
\label{fig:plasma}
\end{figure}

Figure~\ref{fig:plasma} shows the plasma density $n_\pm(z)$ and temperature $\Teff(z)$ established in three simulations with $\zeta=0.2$ and $\lB=10^5$, $10^6$, $10^7$, all with the Maxwell ansatz of the $e^\pm$ distribution function. These values of $\lB$ correspond to the magnetic fields $B=5.6\times 10^8$\,G,  $1.8\times 10^9$\,G, and  $5.6\times 10^9$\,G. One can see that the opaque $e^\pm$ plasma is sustained inside and around the main heating region $|z|\simlt H$. An $e^\pm$ coat is generally expected around compact X-ray sources, since photons can collide and convert to pairs outside the source \citep{Beloborodov99,Beloborodov17c}. The optical depth of the $e^\pm$ plasma is given by
\beq
   \tauT=\int_0^{H_{\rm box}} \sT n_\pm dz.
\eeq
We find $\tauT\approx 4.5$, 6.6, and 10 in the models with $\lB=10^5$, $10^6$, and $10^7$, respectively. Higher $\lB>10^7$ would give $\tauT>10$. Then the diffusion of radiation will become slow compared to its advection by the plasma (\Sect~\ref{advection}). Calculations of burst emission in this regime are deferred to a future work.

In agreement with the estimates in \Sect~\ref{pairs}, $k\Teff$ shown Figure~\ref{fig:plasma} stays in the narrow range of $30-50$\,keV. It is above the Compton temperature $\TC$, as required by the energy balance condition (see \Eq~(\ref{eq:CbalanceT}) for an approximate form of the energy balance). At high altitudes $z$, where heating $\Qdiss$ is negligible, $\Teff$ approaches $\TC$. Curiously, $\Teff(z)$ is not monotonic and its approach to $\TC$ is  slower than might be expected from the exponentially falling $\Qdiss(z)$. This happens because the Comptonization temperature $\Teff(z)$ is shaped not only by the local heating rate, but also by the local densities of photons and $e^\pm$. In particular, the high photon density at the center $z=0$ makes the local Compton cooling efficient, reducing $\Teff$. This causes a reduction in the pair creation rate $\dngg$ toward $z=0$ despite  heating being strongest at $z=0$.

\begin{figure}[t]
\vspace*{-0.6cm}
    \centering
  \includegraphics[width=0.45\textwidth]{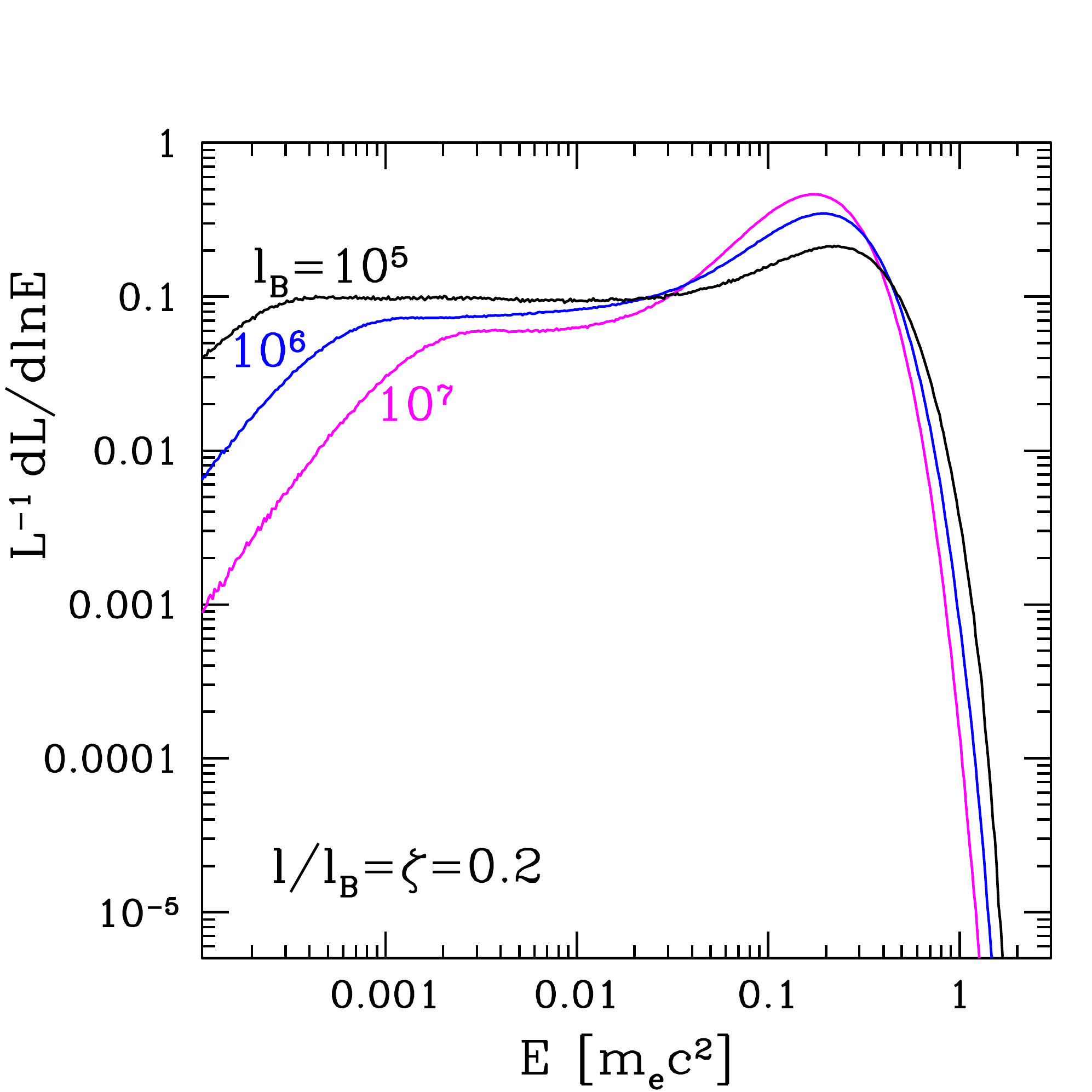}
\caption{Escaping radiation spectrum for the three models shown in Figure~\ref{fig:plasma}.
}
\label{fig:sp_Maxw}
\end{figure}

Figure~\ref{fig:sp_Maxw} shows the escaping radiation spectrum for the three models with Maxwellian plasma, which have $\lB=10^5$, $10^6$, $10^7$ and $\l=0.2\lB$. 
The numerical results are in agreement with the analytical estimates presented in \Sect~\ref{radiation}. In particular, one can see how the slow decrease of $\Teff$ with increasing $\l$ (\Eq~\ref{eq:Teff_M}) results in the slight shift of the spectral peak to lower energies. The photon balance condition enforces the growth of optical depth $\tauT$ in response to the lower $\Teff$, not far from the estimate $\tauT\propto\Teff^{-3}$ in \Eq~(\ref{eq:T-tau}). As a result, the ratio $\tdiff/\tW$ grows (\Eq~(\ref{eq:tW_tdiff})), and so Comptonization becomes more efficient with increasing compactness. This  gives the Comptonized radiation a more pronounced spectral peak, as the Wien peak begins to emerge at $E\sim 3k\Teff$  in the high-compactness models. However, as expected, Comptonization remains far from being saturated. At the high-energy end, $E\gg\Epk$, radiation approaches a Wien spectrum with temperature $\Teff$. At the low-energy end, $E<E_0$, the spectrum follows a Planck distribution. The break position $E_0$ is proportional to $B\propto \lB^{1/2}$; it shifts from $\ep_0\approx 3\times 10^{-4}$ at $\lB=10^5$ to $\ep_0\approx 3\times 10^{-3}$ at $\lB=10^7$.

\begin{figure}[t]
\vspace*{-0.6cm}
    \centering
  \includegraphics[width=0.45\textwidth]{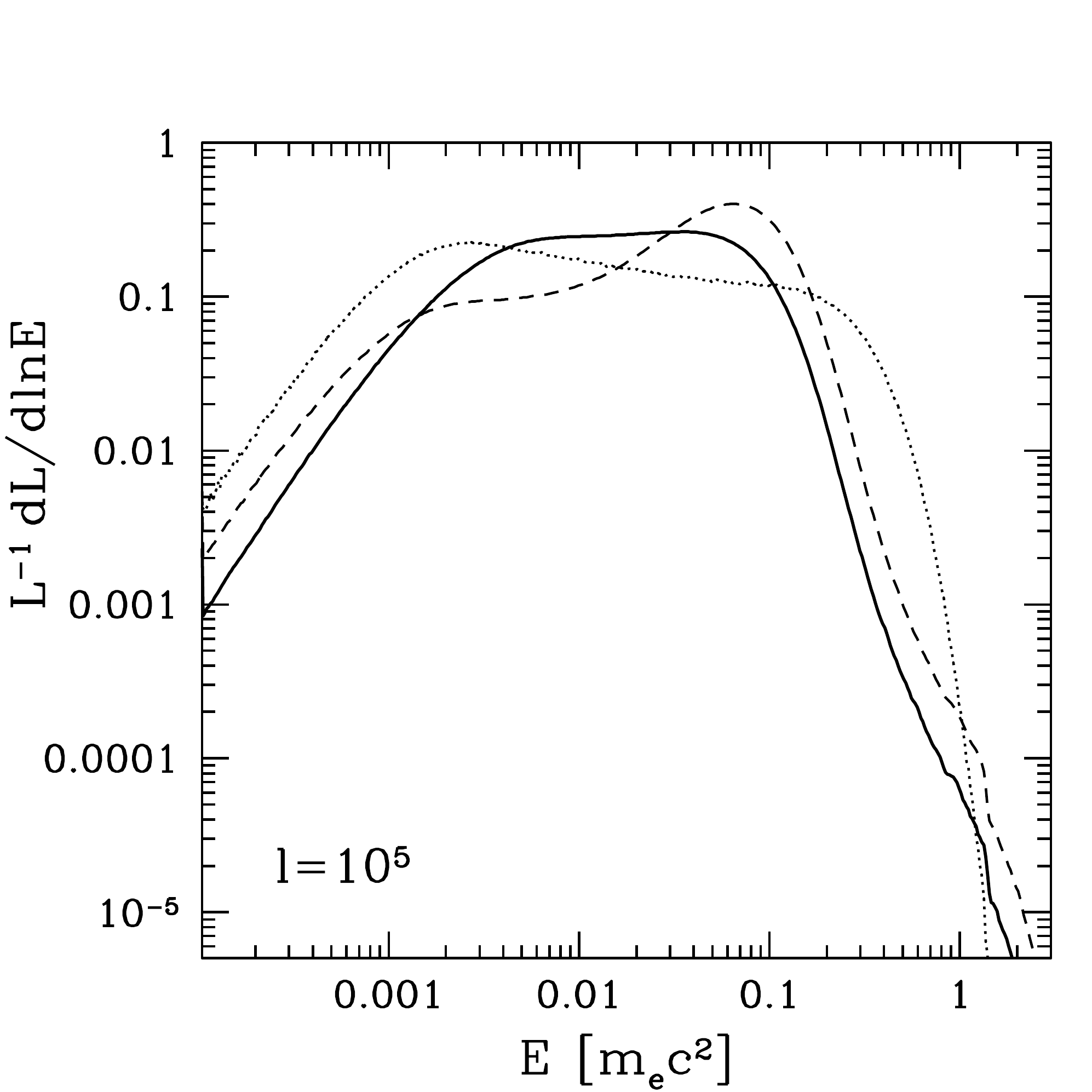}
\caption{
Escaping radiation spectrum of three bursts with equal $\l=10^5$ assuming  different dissipation models: Maxwellian plasma and $\zeta=0.01$ (dotted), nonthermal plasma and $\zeta=0.01$ (dashed), and nonthermal plasma and $\zeta=0.001$ (solid). The nonthermal plasma has an $e^\pm$ distribution described by \Eq~(\ref{eq:fe}) with $\alpha=5$. 
}
\label{fig:sp_alpha}
\end{figure}

Figure~\ref{fig:sp_alpha} shows the burst spectra for a few models with a nonthermal distribution function $f_e(u)$ (\Eq~\ref{eq:fe}) with the high-energy slope $\alpha=5$. This choice of $\alpha$ corresponds to a signifcant fraction $\sim 2\Theff$ of the released power being deposited into relativistic particles $u>1$ (\Eq~\ref{eq:U1}). The  nonthermal plasma more efficiently upscatters photons to $E>m_ec^2$ that create $e^\pm$ pairs. Therefore, when run with the same parameters $\lB$ and $\l$ as the corresponding Maxwellian models, the simulation gives a higher optical depth $\tauT$. Since our simulations neglect the advection of radiation (\Sect~\ref{advection}), we limit the calculations to the regime  of $\tauT\simlt 10$. It can occur in a bright burst with nonthermal plasma only if $\lB\ll\l$, i.e. a small fraction $\zeta\ll 1$ of the magnetic energy is dissipated in the burst. Figure~\ref{fig:sp_alpha} shows the effect of decreasing $\zeta$ on the emitted X-ray spectrum while keeping a constant released power ($\l=10^5=const$). 

Figure~\ref{fig:sp_alpha} also compares the nonthermal and Maxwellian models. Note that the nonthermal plasma gives spectra with an exponential cutoff at a lower $\Epk$. This occurs because the burst self-regulates to a lower $\Teff$, as the nonthermal particles more efficiently upscatter photons to $E>m_ec^2$ and sustain $e^\pm$ creation at a lower $\Teff$.  One can see the nonthermal Comptonized component emerging at $E< m_ec^2$ and extending to $E>m_ec^2$, where it becomes strongly absorbed before reaching $2m_ec^2$. By contrast, no absorption feature is seen in the Maxwellian model; in this case, Comptonization and annihilation emission balance photon-photon absorption while sustaining the Wien (exponential) tail of the spectrum with temperature $\Teff$. 


\section{Discussion}

\subsection{Summary of the emission model}

The radiative mechanism investigated in this paper is generic for fast dissipation events in the outer magnetosphere of a neutron star. The produced X-ray burst has a large dimensionless parameter --- the  compactness $\l\sim 10^4-10^6$ (\Eq~\ref{eq:l}). It scales linearly with the burst power, and is a fraction of the magnetic compactness $\lB$ (\Eq~\ref{eq:lB}), which scales quadratically with the magnetic field $B$. How the magnetic flare is triggered weakly affects its radiation. Therefore, the results should apply equally well to magnetar bursts and to magnetic flares in a tight neutron star binary (Figure~1). 

A key difference of the flare mechanism from existing kinetic plasma simulations of magnetic reconnection is that the plasma density $n_\pm$ is not a given parameter. Instead, the plasma is created during dissipation, and its density is regulated by reactions $\gamma+\gamma\leftrightarrow e^++e^-$. The pre-flare density $n_\pm$ is negligible, and has no effect on the observed emission. 

We have developed a new Monte-Carlo code \texttt{CompPair} for this problem, and simulated bursts produced in magnetic fields $B\sim 10^8-10^{10}\,$G. For a prescribed heating, the code calculates time-dependent radiative transfer with self-consistent $e^\pm$ creation. It shows how the dissipation region becomes dressed in an $e^\pm$ coat, how the $e^\pm$ plasma produces soft photons (via cyclo-synchrotron emission) and Comptonizes the photons to the hard X-ray band before they escape. 

The simulation results are consistent with analytical estimates given in \Sect~3. Most of the burst emission is produced during the quasi-steady phase $t\gg \s/c$, where $\s/c$ is the light crossing time of the dissipation region. Then the nonlinear state of the system is settled by three conditions: energy balance, photon balance, and annihilation balance. The pair density $n_\pm$ settles to values much lower than the photon number density $\nph$: the ratio $n_\pm/\nph$ varies around $10^{-5}$ in the calculated models. This is sufficient to sustain an $e^\pm$ coat; its optical depth $\tauT=5-10$ varies slowly with $\l$. The optically thick $e^\pm$ plasma efficiently upscatters photons to energies $E\sim 0.1 m_ec^2$, with a small fraction of photons ($\sim 10^{-5}$) reaching $E> m_ec^2$ and converting to $e^\pm$. The escaping radiation spectrum is shaped by radiative transfer through the $e^\pm$ coat. 

All the simulated bursts display spectra with an exponential cutoff at energy $\Epk$ in the hard X-ray band. The highest $\Epk$ is achieved when Comptonization occurs with a pure Maxwellian distribution. Then, $\Epk\approx 2.5\, k\Teff \approx 100\,$keV stays almost constant with $\l$ (Figure~\ref{fig:sp_Maxw}). This is an example  of the known ``thermostat'' effect of $e^\pm$ creation \citep{Svensson84}, which keeps the Comptonization temperature $\Teff$ in a narrow range of $30-40\,$keV. An increase of $\Teff$ would generate exponentially more $e^\pm$, and hence more photons, which  immediately cool the plasma. A  decrease in $\Teff$ would suppress $e^\pm$ creation and photon production, leading to a rise in temperature. 

This self-regulation also occurs in bursts with a nonthermal plasma, which sustain a power-law tail of relativistic $e^\pm$. These bursts self-regulate to a lower $\Teff$ of $10-30\,$keV, because nonthermal Comptonization is more efficient in upscattering photons to $E>m_ec^2$ and creating $e^\pm$ pairs.

The overall spectrum shape  shows moderate variations between the calculated models. A typical photon index $\Gamma_{\rm ph}=d\ln N/d\ln E$ at $E<\Epk$ is between $-0.5$ and $-1$. Soft spectra $\Gamma_{\rm ph}<-1$ are found in models where a very small fraction of the magnetic energy is dissipated, $\zeta\simlt 10^{-3}$.

\subsection{Implications for magnetars and mergers}

Our results provide a possible explanation for the observed spectra of magnetar bursts and can shed some light on the dissipation mechanism.
The typical exponential cutoff in observed spectra, $\Epk=20-40$\,keV,  corresponds to low $k\Teff/m_ec^2=0.015-0.03$. The low Comptonization temperatures occur when dissipation generates an $e^\pm$ distribution with a nonthermal tail. Such distributions are expected to occur in magnetic flares, as discussed in \Sect~2.  In our sample nonthermal models, the relativistic $e^\pm$ tail ($u>1$) received $\sim 4$\% of the dissipation power.

We find that the most typical spectra of magnetar bursts are produced when dissipation occurs in stronger magnetic fields $B\sim 10^{10}\,$G with lower dissipation fractions $\zeta \sim 10^{-2}-10^{-3}$ (Figure~\ref{fig:sp_alpha}).
Such fields are found at radii $R\sim 2\times 10^7\,$cm.
This result may be interpreted as follows. If the dissipation is powered by magnetic reconnection, the low $\zeta$ implies a strong guide field, so that the reconnecting field component is $\delta B\sim 0.1 B_{\rm gd}$. The small $\zeta$ may also be consistent with dissipation of Alfv\'en waves of moderate amplitudes $\delta B/B\sim 0.1$, which do not cause a magnetic flare with a global re-structuring of the outer magnetosphere. Such dissipation events are not expected to produce ejecta from the magnetosphere. Thus, in a typical burst, the dissipation region likely remains confined.

The results suggest the following interpretation of the hard X-ray burst of SGR~1935+2154 on 2020 April~28, which was  accompanied by FRB~200428. Its hard spectrum is consistent with a high dissipation fraction $\zeta\simgt 0.1$ in a magnetic field $B\sim 10^{9}\,$G, as expected in a major reconnection event with $\delta B\sim B$ (Figure~1).
 The observed high $\Epk\sim 100\,$keV indicates that the nonthermal $e^\pm$ component was weaker than in typical magnetar bursts.
 When parameterized with index $\alpha$ (\Eq~\ref{eq:fe}), it may correspond to $\alpha\simgt 6$. 

Such magnetic flares should eject plasmoids from the magnetosphere and drive a large-scale magnetic explosion into the magnetar wind \citep{Yuan20}. The blast waves from magnetic flares are expected to produce coherent radio emission in the wind, providing a possible mechanism for repeating FRBs \citep{Beloborodov17b}; \cite{Yuan20} argued that FRB~200428 could be produced by this mechanism. Future detections of FRBs from magnetars could clarify if they are always associated with spectrally hard X-ray bursts, further testing the connection with ejecta-producing magnetic flares in the outer magnetosphere.

Our calculations also suggest what X-ray precursors of neutron star mergers could look like. The precursor luminosity depends on the magnetic fields of the neutron stars, and its spectrum should always extend to $\Epk\sim 0.1 m_ec^2$. The predicted emission is detectable with sensitive X-ray detectors  (for nearby mergers) if the neutron stars have surface magnetic fields $B_\star\sim 10^{13}$\,G. Such binaries are capable of generating luminosities $L\simgt 10^{42}\,$erg/s in the interaction region $R\sim 10^7$\,cm, where $B\sim 10^{10}\,$G. The high luminosities of the precursor flares correspond to large compactness parameters $\l\simgt 10^6$ and $\lB\simgt 10^7$. The energy output of such flares should peak at $E\sim \Epk$ and have an exponential cutoff at $E>\Epk$.

\subsection{Extensions of the model}

The presented simulations used an idealized picture of the dissipation process.   
The magnetic energy was assumed to be uniform in the simulation box while in a real magnetospheric burst there are spatial and temporal variations of the reconnecting magnetic field. Then, effectively, one may observe a superposition of spectra with varying $\l$ and $\lB$. However, the described self-regulation process across a broad parameter space should prevent dramatic spectral changes. Note also that the models presented in Figures~\ref{fig:plasma}-\ref{fig:sp_alpha} were calculated with a Gaussian distribution of the heating rate in the dissipation layer, $\Qdiss(z)=\Qdiss^0 \exp(-z^2/2\h^2)$, where $\h/\s=\vrec/c=0.1$, a canonical thickness of a reconnection layer. Different $\h$ is possible and may deserve further investigation. We have run a few models with a smaller $\h=0.03\s$ and saw minor changes in results. 

Two additional effects may influence future models of magnetic reconnection around neutron stars:
\medskip
\\
(1) The reconnection rate may be affected by the radiation pressure developing around the dissipation layer. The $e^\pm$ coat implies a nominal  magnetization parameter $\sigma_0\sim 0.2\lB/\tauT$ (\Eq~\ref{eq:sigma0}), which is huge. However, the $e^\pm$ mass is dominated by the effective inertial mass of radiation interacting with the plasma, $U_{\rm rad}/c^2\gg n_\pm m_e$. One can define the effective magnetization as $\sigma=B^2/4\pi (U_{\rm rad}+n_\pm m_ec^2)$, and it is not much greater than unity. This implies that radiation pressure may become competitive with magnetic stresses that drive the reconnection process.
\medskip
\\
(2) The other effect was discussed in \Sect~\ref{advection}: the magnetic field lines moving through the reconnection region advect the $e^\pm$-photon fluid.
When $\tauT>c/v_{\rm rec}$, radiation diffusion becomes slower than advection. Then a large part of the produced radiation will be advected sideways along the reconnection layer, and ejected together with the magnetic plasmoids. This effect was not modeled in our simulations, which were limited to the diffusion-dominated regime of $\tauT\simlt c/\vrec$.
\medskip

Plasmoids ejected by the magnetic flare away from the star will immediately expand and release their radiation, so their effective $\sigma$ quickly rises from $B^2/4\pi U_{\rm rad}$ to $B^2/4\pi n_\pm m_e$. The final $\sigma$ is comparable to $\sigma_0$ inside the reconnection layer, because the escaping plasmoid density $n_\pm$ is only moderately reduced by $e^\pm$ annihilation.\footnote{$e^\pm$ annihilation becomes slower than expansion when the ejecta Thomson optical depth drops to $\sim 1$, which is not far below $\tauT$ inside the magnetic flare. Note that flares in the outer magnetosphere considered here are different from magnetar giant flares. Giant flares  produce ejecta with enormous initial $\tauT$, and their $n_\pm$ is strongly reduced by annihilation \citep{Beloborodov20}.}
The resulting magnetization parameter of the flare ejecta may be estimated as
\beq
   \sigma_{\rm ej}\sim 0.1\,\lB.
\eeq
The high $\sigma_{\rm ej}$ implies that the ejecta will accelerate to a high Lorentz factor as it leaves the magnetosphere, launching an ultra-relativistic blast wave into the surrounding wind from the neutron star (or the wind from the neutron star binary in the pre-merger systems). The ejecta acceleration was observed in the simulation of \cite{Yuan20}, which assumed $\sigma\rightarrow\infty$ (the force-free limit of magnetohydrodynamics). The ultra-relativistic acceleration of the flare ejecta is essential for the blast wave scenario of FRB production \citep{Beloborodov20}.

Note that our calculations neglected the effect of the magnetic field on the Compton scattering cross section. This approximation is good for dissipation events in the outer magnetosphere, as photons of interest have energies $E\gg \hbar eB/m_ec^2\sim 0.1\, B_{10}$\,keV. For more powerful magnetic flares, with stronger $B$, it becomes important to follow the transfer of photons in two linear polarization states, which have different scattering cross sections in the ultrastrong $B$. 

Ultrastrong flares would also develop much higher $e^\pm$ densities, and radiation trapped in the $e^\pm$ plasma would become thermalized. This is the situation of giant flares discussed by \cite{Thompson96}. An estimate for the thermalization transition may be obtained by comparing the plasma temperature predicted by our models, $T\sim 2\times 10^8$\,K, with the characteristic blackbody temperature of the dissipated magnetic energy, $T_{\rm BB}\sim (U_B/a)^{1/4}\approx 1.5\times 10^8\,B_{10}^{1/2}$\,K, where $a=7.56\times 10^{-15}$\,erg\;cm$^{-3}$\,K$^{-4}$ is the radiation constant. This comparison shows that thermalization must occur in flares with $B>B_{\rm th}\sim 3\times 10^{10}$\,G. The plasma temperature in flares with $B\ll B_{\rm th}$ very slowly decreases with $B$, reaching a minimum of $kT\sim 10-20$\,keV at $B\sim B_{\rm th}$, and then grows as $B^{1/2}$ at $B>B_{\rm th}$. At the QED field  $B_{\rm Q}=m_e^2c^3/\hbar e=4.44\times 10^{13}\,$G, the flare temperature approaches $kT_{\rm BB}\sim m_ec^2$, which gives $n_\pm\sim \nph$ and a huge optical depth. The spectrum radiated from the photosphere of this hot fireball was discussed by \cite{Lyubarsky02}.

\medskip

This work was supported by NASA grant NNX~17AK37G, NSF grant AST~2009453, Simons Foundation grant \#446228, and the Humboldt Foundation.


 \bibliography{bursts}

 \end{document}